\documentclass[pra,amsmath,amssymb,twocolumn,superscriptaddress,showpacs]{revtex4}

\usepackage{amsmath,amssymb}
\usepackage[usenames]{color}
\usepackage{amssymb}
\usepackage{grffile}
\usepackage[pdftex]{graphicx}
\usepackage{amsmath, amstext, amssymb, amsfonts, amsxtra}
\usepackage{textcomp}
\usepackage{xspace}
\usepackage{epstopdf}
\DeclareGraphicsExtensions{.eps}
\usepackage[percent]{overpic}
\usepackage{enumerate}

\newcommand{\ket}[1]{|#1\rangle}

\newcommand{\bra}[1]{\langle#1|}

\def \ell{{d}}

\newcommand{\sutd}{Singapore University of Technology and Design, 8 Somapah Road 487372, Singapore}
\newcommand{\hannover}{Institut f\"ur Theoretische Physik, Leibniz Universit\"at Hannover, Appelstr. 2, DE-30167 Hannover, Germany}
\newcommand{\madrid}{Instituto de Estructura de la Materia, CSIC, Serrano 123, E-28006 Madrid, Spain}
\newcommand{\stuttgart}{Institut fur Theoretische Physik III, Universit\"at Stuttgart, Pfaffenwaldring 57, 70550 Stuttgart, Germany}
\newcommand{\majulab}{MajuLab, CNRS-UNS-NUS-NTU International Joint Research Unit, UMI 3654, Singapore}

\begin{document}

\title{Density-dependent synthetic magnetism for ultracold atoms in optical lattices}                     

\author{Sebastian Greschner} 
\affiliation{\hannover}
\author{Daniel Huerga} 
\affiliation{\madrid}
\affiliation{\stuttgart}
\author{Gaoyong Sun} 
\affiliation{\hannover}
\author{Dario Poletti}
\affiliation{\sutd}
\affiliation{\majulab}
\author{Luis Santos}
\affiliation{\hannover}

\begin{abstract}
Raman-assisted hopping can allow for the creation of 
density-dependent synthetic magnetism for cold neutral gases in optical lattices. We show that 
the density-dependent fields lead to a non-trivial interplay between density modulations and chirality. 
This interplay results in a rich physics for atoms in two-leg ladders, characterized by a density-driven Meissner- to vortex-superfluid 
transition, and a non-trivial dependence of the density imbalance between the legs. Density-dependent fields also lead to intriguing physics in square lattices. In particular, it leads to a density-driven transition between a non-chiral and a chiral superfluid, both characterized by non-trivial charge density-wave amplitude.
We finally show how the physics due to the density-dependent fields may be easily probed in experiments 
by monitoring the expansion of doublons and holes in a Mott insulator, which presents a remarkable dependence on quantum fluctuations.
\end{abstract}


\maketitle


%
\begin{figure*}[t]
\begin{center}
\includegraphics[width =\linewidth]{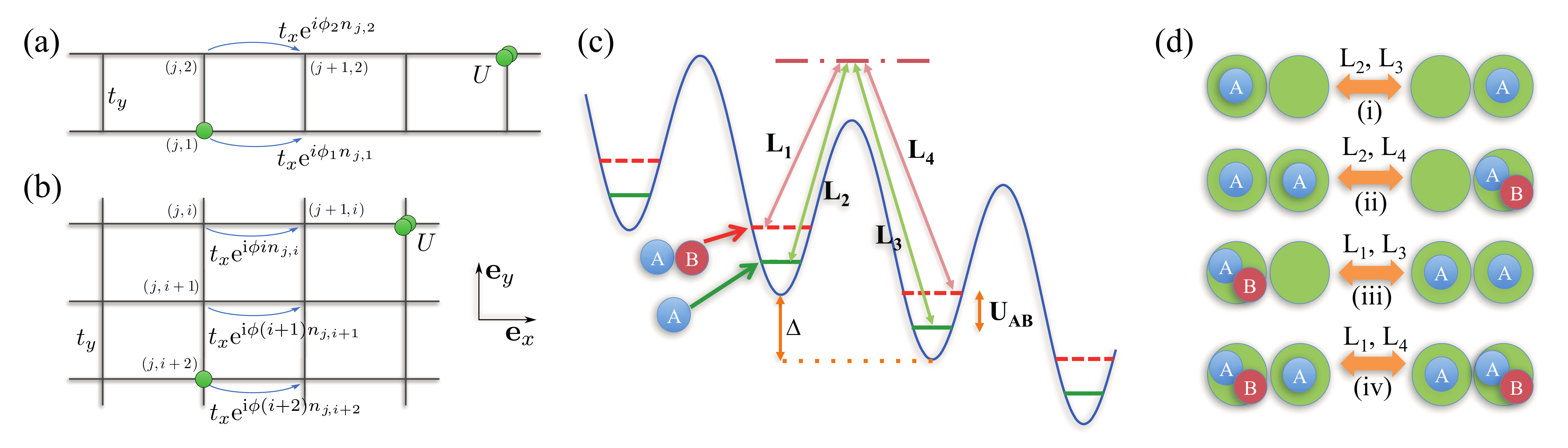}
\caption{(a) Sketch of the density-dependent Peierls phases of the model~\eqref{eq:Heff-2D_gen} on a ladder (see also Eq.~\eqref{eq:Hladder}) and (b) of the 2D-square lattice. (c) Scheme of the creation of a density-dependent Peierls phase using Raman assisted hoppings. (d) Raman assisted hops (i)--(iv) as discussed in the main text. }
\label{fig:1}
\end{center}
\end{figure*}


\section{Introduction}

Orbital magnetism is crucial in condensed-matter physics. In particular, it plays a fundamental role 
in the integer and fractional quantum Hall physics~\cite{VonKlitzing1980,Tsui1982,Laughlin1983}, as well as in related topics 
such as anyons~\cite{Wilczek1982} and  topological insulators~\cite{Kane2005,Bernevig2006}. 
Ultracold gases offer extraordinary possibilities for the controlled experimental simulation of quantum many-body systems~\cite{Bloch2008}. However, 
experiments are typically performed with neutral particles precluding the direct quantum simulation of orbital magnetism. 
Interestingly, synthetic magnetism can be engineered in cold neutral gases, currently constituting 
 a major topic in cold-atom research~\cite{Dalibard2011,Goldman2014}. Proper laser arrangements have been used to induce 
artificial magnetic fields and spin-orbit coupling~\cite{Lin2011,Lin2011b}. In addition, recent experiments have 
demonstrated the creation of synthetic magnetism in 2D optical lattices~\cite{Aidelsburger2013,Miyake2013}, and realized the 
Meissner-superfluid~(MSF) to vortex-superfluid~(VSF) transition~\cite{Orignac2001} with bosons in square optical ladders~\cite{Atala2014,Mancini2015,Stuhl2015}. 

However, in all these experiments the fields created were static since there was no feedback of the atoms on the field. 
Nevertheless, such a dynamical feedback plays an important role 
in various areas of physics, ranging from condensed-matter~\cite{Levin2005} to quantum chromodynamics~\cite{Kogut1983}.
In order to experimentally simulate such dynamical gauge fields in optical lattices various ideas have been recently suggested \cite{Cirac2010,Zohar2011,Kapit2011,Zohar2012,Banerjee2012,Zohar2013,Banerjee2013,Zohar2013b,Tagliacozzo2013}. 

While in those kind of dynamical gauge fields it is crucial to discriminate particle and gauge fields degrees of freedom, 
gauge fields determined by the particle density appear in a variety of problems in condensed-matter physics, including effective field theories for the 
fractional quantum Hall effect~\cite{Zhang1989} and their extension to generalized statistics in one dimension~\cite{Rabello1996}, spin models of quantum magnetism~\cite{Fradkin1989}, and chiral solitons~\cite{Aglietti1996}. Indeed, the atomic back-action on the synthetic gauge field in cold gases experiments is expected to lead to these last type of excitations in Bose-Einstein condensates~\cite{Edmonds2013}. 

From the experimental standpoint, Raman-assisted hopping of cold atoms~\cite{Aidelsburger2013,Miyake2013,Atala2014,Mancini2015,Stuhl2015} can be used in 1D optical lattices to realize occupation-dependent gauge fields that result in effective anyon statistics and thus a clean experimental realization of a 1D {\it anyon}-Hubbard model~\cite{Keilmann2011,Greschner2015AHM}. This model presents a wealth of new physics, including statistically-induced phase transitions~\cite{Keilmann2011}, novel superfluid phases~\cite{Greschner2015AHM}, asymmetric momentum distributions~\cite{Hao2009}, and intriguing dynamics~\cite{DelCampo2008,Hao2012,Wang2014}. 

In this paper we show how a Raman-laser scheme can be employed for the
realization of density-dependent synthetic magnetism (DDSM) in one and two
dimensions and study its effects in ladder and square lattice geometries.
In particular, we are interested in a system described by the following Hamiltonian
\begin{eqnarray}
\mathcal{H}&=&-\sum_{\mathbf r}  \left [ t_x b_{{\mathbf r}+{\mathbf e}_x}^\dag e^{{\rm i}\phi_{\mathbf r} n_{\mathbf r}} b_{{\mathbf r}} 
+t_y  b_{{\mathbf r}+{\mathbf e}_y}^\dag b_{{\mathbf r}}  +\mathrm{H.c.}\right ]  \nonumber \\
&&+ \frac{U}{2}\sum_{{\mathbf r}} n_{{\mathbf r}}(n_{{\mathbf r}}-1) - \mu\sum_{\mathbf{r}} n_\mathbf{r},
\label{eq:Heff-2D_gen}
\end{eqnarray}
where $b_{{\mathbf r}}$ ($b_{{\mathbf r}}^\dagger$) is a bosonic annihilation (creation) operator acting on site $\mathbf{r}=(i,j)$ of the lattice, and $n_{\mathbf r}=b_{{\mathbf r}}^\dagger b_{{\mathbf r}}$ is the number operator. As it will be discussed below, the experimental implementation leads to a three-body hardcore constraint on the onsite occupation, i.e. $n_{\mathbf{r}}=0,1,2$. The first term in ($\ref{eq:Heff-2D_gen}$) accounts for the hopping of bosons along the two directions of the lattice, defined by lattice vectors ${\mathbf e}_x=(1,0)$ and ${\mathbf e}_y=(0,1)$, while the second and third terms account for the usual onsite Hubbard interaction and the chemical potential which fixes the total density of the system, respectively (see Fig.~\ref{fig:1}~(a) and (b)). As it will be shown below, the density dependent Peierls phase of the hopping amplitude $(e^{{\rm i}\phi_{\mathbf r} n_{\mathbf r}})$ can be chosen in such a way that an effective net-magnetic flux per unit-cell is created. In particular, we will concentrate in the case were the phases depend only on the position in the $y$-direction, i.e. $\phi_{\mathbf{r}}=\phi_{j}$. Due to the operator nature of this phase, quantum fluctuations of the density will crucially affect the effective magnetic flux. In this work, we demonstrate that DDSM has important consequences for bosons in two-leg ladders and 2D square lattices, leading to a non-trivial interplay between chirality and density modulations.

The structure of the paper is as follows. 
In Sec.~\ref{sec:DDF} we comment on the realization of DDSM using Raman-assisted hopping.
In Sec.~\ref{sec:DDSMLadders} we analyze the consequences 
of DDSM in optical ladders, whereas in Sec.~\ref{sec:2D} we focus on the case of 2D square lattices. In ladders, this interplay results in a density-driven Meissner-superfluid~(MSF) -- Vortex-superfluid~(VSF) transition with a non-trivial density imbalance between the legs. In square lattices DDSM induces a similar transition between a non-chiral superfluid~(SF) and a chiral superfluid~(CSF), both presenting a non-trivial density-wave amplitude. 
Section~\ref{sec:Dynamics} is devoted to the dynamics of particles and holes, which are crucially affected by the DDSM, as illustrated by the expansion of doublons and holes in a Mott insulator~(MI), which presents an intriguing dependence on quantum fluctuations.
Finally in Sec.~\ref{sec:Summary} we summarize our results.



\section{Density-dependent fields}
\label{sec:DDF}

In this section we propose a possible experimental scheme for the realization of DDSM. First, we briefly review the proposal for the creation of a density-dependent Peierls phase in 
one dimensional lattices, as described in Ref.~\cite{Greschner2015AHM}, which is the key ingredient for the realization of DDSM. In following subsections, we discuss how this scheme naturally extends to higher dimensional lattices and how it may be adjusted to effectively reproduce the density-dependent Peierls phases of Model~\eqref{eq:Heff-2D_gen}.

\subsection{Two-component system}
We consider a bosonic species with two internal states, $|A\rangle$  and $|B\rangle$.  
As shown below for the specific case of $^{87}$Rb, we may choose $|A\rangle\equiv |F=1,m_F=-1\rangle$ and $|B\rangle\equiv |F=2,m_F=-2\rangle$. 
A detailed discussion of other species can be found in the supplemental information of Ref.~\cite{Greschner2015AHM}.
Both components are confined to the lowest band of a tilted 1D optical lattice along the $x$ axis, of spacing $D$, and depth $V_0=sE_R$, with $E_R=\hbar^2 \pi^2/2mD^2$ 
the recoil energy. The Hilbert-space of a single lattice site thus constitutes of empty sites (0), single occupied sites (A) or (B), doubly occupied sites (AA), (BB) or (AB), etc.  Without tilting there is a hopping rate $J$ to nearest neighbors. The lattice tilting induces an energy shift $\Delta$ from site to site as shown in Fig.~\ref{fig:1}~(c). 

We denote as $w(x-j D)$ the Wannier function at site $j$. Due to the tilting, it is convenient 
to use Wannier-Stark states. For $J\ll \Delta$, the Wannier-Stark state centered at site $j$ may be approximated as 
$\psi_{j}(x)\simeq w(x-jD)+\frac{J}{\Delta}[w(x-(j+1)D)-w(x-(j-1)D)]$~\cite{MiyakeThesis}.
The 3D on-site wavefunction at site $j$ is $\Phi_j({\bf r})=\psi_j(x)\varphi(y,z)$, where $\varphi(y,z)$ is given by the strong transversal confinement. For simplicity 
we assume below $\varphi(y,z)\simeq w(y)w(z)$. 

On-site interactions between atoms in states $\alpha$ and $\beta$~(for $\alpha$, $\beta=$ A, B) are 
characterized by the coupling constant $U_{\alpha,\beta}=\frac{4\pi\hbar^2a_{\alpha,\beta}}{m}\int d^3{\mathbf r} |w({\mathbf r})|^4$, with $a_{\alpha,\beta}$ the corresponding scattering length. For a sufficiently deep lattice, the evaluation of the on-site interactions 
is simplified by means of the harmonic approximation~\cite{Bloch2008}: $\Phi({\bf r})\simeq (\sqrt{\pi} l)^{-3/2}e^{-r^2/l^2}$, 
where $l=D s^{-1/4}/\pi$. 
Using this approximation we obtain:
$U_{\alpha,\beta}\simeq \sqrt{2}\pi^{5/2} s^{3/4}\frac{\hbar^2 a_{\alpha,\beta}}{mD^3}$. 
As shown in Ref.~\cite{Greschner2015AHM} the scheme may as well be realized with fermionic species; then inter-species on-site interaction $U_{AB}$ is possible.

\subsection{Raman-assisted hopping}

No direct hopping occurs since  $J\ll \Delta,U_{\alpha,\beta}$. Raman-assisted hopping is realized with the set-up of Fig.~\ref{fig:1}~(c) formed by four 
lasers $L_{j=1,\dots,4}$, with Rabi frequencies $\Omega_j=|\Omega_j|e^{{\rm i}\phi_j}$, wave vectors ${\mathbf k}_j$, and frequencies $\omega_j$.
$L_{1,4}$ have linear polarization and $L_{2,3}$ circular $\sigma_-$ polarization and couple states $|A\rangle$ and $|B\rangle$ far from resonance. 
$|B\rangle$ is just affected by lasers $L_{1,4}$ due to selection rules. Although both $L_{2,3}$ and $L_{1,4}$ couple to $|A\rangle$, the coupling with $L_{1,4}$ 
can be made much smaller than that of $L_{2,3}$ (for a detailed discussion see the supplemental information of Ref.~\cite{Greschner2015AHM}). Hence we may assume below that $|A\rangle$ is just affected by $L_{2,3}$.

Following Ref.~\cite{MiyakeThesis}, we evaluate the Raman-assisted hopping, $J_{nm}$, given by lasers $L_{n=1,2}$ and $L_{m=3,4}$, from site $j$ to site $j+1$:
\begin{equation}
J_{nm}=\frac{V_{nm}}{4}e^{{\rm i}\phi_{nm}}
\int d^3{\mathbf r}\,  \Phi_{j+1}({\mathbf r})^* e^{{\rm i}\delta{\mathrm k}^{nm}\cdot {\mathbf r}} \Phi_{j}({\mathbf r}),
\end{equation}
where $\phi_{nm}=\phi_n-\phi_m$, $\delta{\mathbf k}^{nm}={\mathbf k}_n-{\mathbf k}_m$, and $V_{nm}=\frac{\hbar |\Omega_n||\Omega_m|}{\delta}$, with $\delta$ the (large) detuning to the one-photon transitions. 
For $J\ll \Delta$ and $s\gg 1$, we may approximate:
\begin{equation}
J_{nm}\simeq {\rm i}\left (\frac{V_{nm}}{2\Delta} \right ) J \sin \left ( \frac{\delta k_x^{nm} D}{2} \right ) e^{{\rm i} \delta k_x^{nm} D(j+1/2)}e^{{\rm i}\phi_{nm}}.
\end{equation}
Note that $\delta k_x\neq 0$ is necessary to establish a significant assisted hopping~\cite{MiyakeThesis,Miyake2013,Aidelsburger2013}.
Each laser pair couples a different Raman transition (see Fig.~\ref{fig:1}~(d)):
\begin{enumerate}[(i)]
\item $J_{23}$ characterizes the hopping (A,0)$\to$(0,A), which is accompanied by an energy shift $\Delta E=-\Delta$. We hence demand $\omega_2-\omega_3=-\Delta$ and the transition amplitude is given by $V_{23}\simeq \frac{1}{2}\frac{\Omega_2\Omega_3^*}{\delta}$ including the appropriate Clebsch-Gordan coefficients for the specific case of $^{87}$Rb.
\item (A,A)$\to$(0,AB) is given by $J_{24}$, being characterized by $\Delta E=-\Delta+U_{AB}$;  we impose $\omega_2-\omega_4= -\Delta+U_{AB}+U$, with $U\ll U_{AB},\Delta$ and the amplitude $V_{24}=\frac{1}{\sqrt{6}}\frac{\Omega_2\Omega_4^*}{\delta}$
\item $J_{13}$  is linked to the hop (AB,0)$\to$(A,A); the energy shift is $\Delta E=-\Delta-U_{AB}$;  we demand $\omega_1-\omega_3\simeq -\delta-U_{AB}-U$. $V_{13}=\frac{1}{\sqrt{6}}\frac{\Omega_1\Omega_3^*}{\delta}$.
\item (AB,A)$\to$(A,AB) is given by $J_{14}$; the energy shift is $\Delta E=-\Delta$;  we impose $\omega_1-\omega_4= -\Delta$. The transition amplitude is given by $V_{14}=\frac{1}{3}\frac{\Omega_1\Omega_4^*}{\delta}$.
\end{enumerate}
The frequencies $\omega_j$ are chosen such that they compensate the lattice tilting, and hence no Bloch oscillation is induced in the rotating frame.
In this frame process (ii) is accompanied by an energy shift $U$, (iii) by a shift $-U$, and (i) and (iv) have no associated shift. 
$U$ may be hence understood as an effective on-site interaction energy. Alternatively, these energy shifts are compatible with an on-site interaction $U_{AB}$, and an effective nearest-neighbor interaction  $V=U_{AB}-U$. We will return to this point when discussing the extension to 2D lattices.

Note that processes (i) and (iv) are energetically degenerated, but they may be addressed with different lasers due to selection rules. This point constitutes the major 
drawback of the proposal of Ref.~\cite{Keilmann2011}. In that proposal, a single component, A, was considered, and process (iv) was of the form (AA,A)$\to$(A,AA), which cannot be resolved 
from process (i). As a result, in the scheme of Ref.~\cite{Keilmann2011}, both the combination of $L_2$ and $L_3$, and of $L_1$ and $L_4$ address both (i) and (iv), preventing the realization of the desired density-dependent Peierls phase. 
The two processes may be just discerned by considering a very small detuning $\delta<U_{AA},\Delta$, which would be accompanied by very large spontaneous emission losses.

\subsection{Spurious processes}

Undesired spurious processes are in principle possible:
\begin{enumerate}[(i)]
\setcounter{enumi}{4}
\item (A,0) $\to$ (0,B); $\Delta E=-\Delta$
\item (A,A) $\to$ (0,AA): $\Delta E=-\Delta+U_{AA}$
\item (AA,0) $\to$ (A,A): $\Delta E=-\Delta-U_{AA}$
\item (AB,A) $\to$ (B,AA): $\Delta E=-\Delta+\delta U$, with $\delta U=(U_{AA}-U_{AB})$
\item (AA,B) $\to$ (A,AB); $\Delta E=-\Delta-\delta U$
\end{enumerate}
Process (v) is just possible with $J_{24}$ or $J_{13}$. But these laser combinations are (quasi-)resonant with $-\Delta\pm U_{AB}$. 
For $U_{AB}\gg W$, with $W$ the width of the Raman resonance~(typically of the order of  $50$ Hz~\cite{MiyakeThesis}), 
process (v) is far from resonance with either  $J_{24}$ or $J_{13}$. 
To neglect the (vi) and (vii) processes one needs $U_{AA}\gg W$. 
In contrast, to avoid (viii) and (ix) one must demand $\delta U \gg W$. The latter condition is certainly more strict, but may be attained in experiments as shown in the supplemental information of Ref.~\cite{Greschner2015AHM}.

\subsection{Effective 1D Hamiltonian}

We assume  $\frac{|\Omega_1||\Omega_4|}{4}=\frac{|\Omega_2||\Omega_3|}{3}=
\frac{|\Omega_1||\Omega_3|}{2\sqrt{3}}=\frac{|\Omega_2||\Omega_4|}{2\sqrt{3}}=\Omega^2$, 
$\Omega_1=|\Omega_1|e^{-{\rm i}\phi}$, and $\Omega_{j=2,3,4}=|\Omega_j|$
and obtain the transition amplitudes
$V_{23}\simeq\frac{\Omega^2}{\delta}$, $V_{24}=\sqrt{2}\frac{\Omega^2}{\delta}$, 
$V_{13}=\sqrt{2}\frac{\Omega^2}{\delta}e^{-{\rm i}\phi}$, and 
$V_{14}=2\frac{\Omega^2}{\delta}e^{-{\rm i}\phi}$.
Note that an additional factor $\sqrt{2}$ is used to mimic bosonic enhancement. 
We denote as $c_j$ the bosonic operator corresponding to the Fock-state manifold $\{ |0\rangle$, $|1\rangle\equiv |A\rangle$, $|2\rangle\equiv |AB\rangle\}$. 
Assuming ${\mathbf k}_{1,2}=k {\mathbf e}_y$, and ${\mathbf k}_{3,4}=k {\mathbf e}_x$, and $kD=\pi$, then 
\begin{equation}
\mathcal{H}=-t \sum_j  (-1)^j \left [ c_{j}^\dag e^{{\rm i}\phi n_j} c_{j+1} +\mathrm{H.c.}\right ] + \frac{U}{2}\sum_j n_j(n_j-1)
\label{eq:Heff-0}
\end{equation}
with $n_j=c_j^\dag c_j$, and $t=\left (\frac{\Omega^2/\delta}{2\Delta} \right ) J$. 
Typical values of the Raman-assisted hopping rate, $t$, are of the order of few tens of Hz~\cite{MiyakeThesis}.
Note that the factor $(-1)^j$, which results from the $x$ projection of $\delta {\mathbf k}$, may be easily eliminated by redefining the bosonic operators in the form: 
$b_{4l}=c_{4l}$, $b_{4l+1}=c_{4l+1}$, $b_{4l+2}=-c_{4l+2}$, $b_{4l+3}=-c_{4l+3}$, with $l$ an integer. In this way we obtain the 1D model:
\begin{equation}
\mathcal{H}=-t \sum_j \left [ b_{j}^\dag e^{{\rm i}\phi n_j} b_{j+1} +\mathrm{H.c.} \right ] + \frac{U}{2}\sum_j n_j(n_j-1)
\label{eq:Heff}
\end{equation}

\subsection{Density-dependent gauge fields in 2D lattices}

For a 2D square lattice or ladder, one may proceed as in Refs.~\cite{Miyake2013,Aidelsburger2013}, assuming assisted hopping along $x$, 
and natural hopping along $y$. This is however problematic, as one can clearly understand from the 
alternative picture mentioned above (section II~B), in which the on-site interactions remain characterized by $U_{AB}$, but an effective nearest-neighbor interaction, $V$, is induced along $x$. 
In contrast, along $y$ there is no nearest-neighbor interaction. Although this asymmetric extended-Hubbard model may have interest in itself, it is not the model to be explored in this work.

An effective model with only on-site interactions and a density-dependent gauge demands both directions to be Raman-assisted.
Following the same arguments as above, we evaluate the 
assisted hopping given by lasers $n$ and $m$ from a site ${\mathbf r}=(D_x r_x,D_y r_y)$ to the site ${\mathbf r}+D_j {\mathbf e}_j$, with ${\mathbf e}_{j}$ the 
unit vector along the $j=x,y$ direction, and $D_j$ the lattice spacing along that direction:
\begin{equation}
J_{nm}^{(j)}\simeq \left (\frac{{\rm i}V_{nm}J_j}{2\Delta_j} \right )e^{{\rm i}\phi_{nm}} \sin \left ( \frac{\delta k_j^{nm} D_j}{2} \right ) e^{{\rm i} \mathbf {\delta k}^{nm} \cdot  ({\mathbf r}+D_j\frac{{\mathbf e}_j}{2})},
\end{equation}
where $\Delta_j$ and $J_j$ are, respectively, the tilting and the hopping without tilting along the $j$ direction.

\begin{figure*}[t]
\centering
\includegraphics[width=1\linewidth]{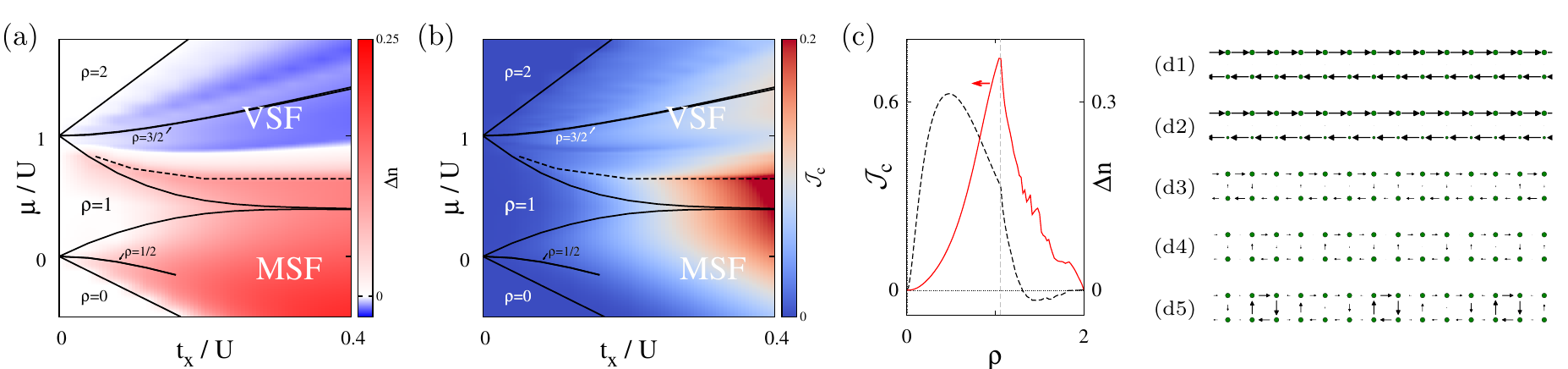}
 \caption{~(a) and (b) Phase diagram for a ladder with $t_y=t_x$, $\phi_1=0.8\pi$, and $\phi_2=0$ as function of $t_x/U$ and chemical potential $\mu$ computed with the density matrix renormalization group~(DMRG). The color code indicates (a) the particle density imbalance between the legs $\Delta n$ and (b) the chiral current $\mathcal{J}_c$ (obtained from simulations with $L=24$ rungs). 
 Solid lines mark the MI with $\rho=1$, and~(very narrow) with $\rho=1/2$ and $3/2$ (extrapolated to the thermodynamic limit from systems with up to $L=96$ rungs). The dashed line denotes the MSF-VSF transition. 
 (c) $\Delta n$~(dashed) and $\mathcal{J}_c$~(solid) for the same parameters as in (a) and (b) but $U=0$ and $L=48$.
(d1-5) Typical particle density and current configurations for $U=0$ and (d1) $\rho=0.1$, (d2) $\rho=0.63$, (d3) $\rho=1.25$, (d4) $\rho=1.46$, (d5) $\rho=1.77$. The size of the circles is proportional to the onsite-density, he arrows encode the strength of the local currents.
  \label{fig:2}}
\end{figure*}

\subsection{Four-laser arrangement}

We first consider the same arrangement of four Raman-lasers as discussed above. We assume $\Delta_x=\Delta_y=\Delta$, and $J_x=J_y=J$.
For ${\mathbf k}_3={\mathbf k}_4=\frac{\pi}{D_x}{\mathbf e}_x$, ${\mathbf k}_2=\frac{\pi}{D_y}{\mathbf e}_y$, and 
${\mathbf k}_1=\frac{\pi+\phi}{D_y}{\mathbf e}_y$:
\begin{equation}
J_{nm}^{(j=x,y)}({\mathbf r})= \left (\frac{{\rm i}V_{nm}J_j}{2\Delta_j} \right )e^{{\rm i}\phi_{nm}}e^{{\rm i}\pi(r_x+r_y+1/2)} \lambda_{nm}^{(j)},
\end{equation}
with $\lambda_{23,24}^{(x,y)}=1$, and $\lambda_{13,14}^{(x)}=e^{{\rm i}\phi j_y}$, $\lambda_{13,14}^{(y)}=e^{{\rm i}\phi (j_y+1/2)} \cos(\phi/2)$.
Assuming $\Omega_{2,3}=\Omega$, $\Omega_{1,4}=\sqrt{2}\Omega$, $\phi_{1,\dots,4}=0$, we obtain the effective 
Hamiltonian:
\begin{eqnarray}
\mathcal{H}&=&-t\sum_{\mathbf r}  (-1)^{r_x+r_y} \sum_{j=x,y} \left [ c_{{\mathbf r}+{\mathbf e}_j}^\dag f_j(r_y,n_{\mathbf r}) c_{{\mathbf r}} +\mathrm{H.c.}\right ]  \nonumber \\
&+& \frac{U}{2}\sum_{{\mathbf r}} n_{{\mathbf r}}(n_{{\mathbf r}}-1).
\label{eq:Heff-2D_sign}
\end{eqnarray}
with $f_x(r_y,n_{\mathbf r})=e^{{\rm i}\phi r_y n_{\mathbf r}}$, and $f_y(r_y,n_{\mathbf r})= \cos(\phi n_{{\mathbf r}})  e^{{\rm i}\phi (r_y+1/2) n_{\mathbf r}}$.
As for the 1D problem, the factors $(-1)^{j_x+j_y}$ originate from the projection of all $\mathbf{\delta k}^{nm}$ along $x$ and $y$ in order to achieve 
assisted hopping along both directions. They may be readily eliminated by introducing the transformation $b_{4 n_x+s_x,4n_y+s_y}=\epsilon(s_x,s_y)c_{4 n_x+s_x,4n_y+s_y}$, 
where $\epsilon(s_x,s_y)=-1$ if $ (s_x+s_y)\mod 4 >1$ and  $\epsilon(s_x,s_y)=1$ otherwise. With this transformation we obtain:
\begin{eqnarray}
\mathcal{H}&=&-t\sum_{\mathbf r} \sum_{j=x,y} \left [ b_{{\mathbf r}+{\mathbf e}_j}^\dag f_j(r_y,n_{\mathbf r}) b_{{\mathbf r}} +\mathrm{H.c.}\right ]  \nonumber \\
&+& \frac{U}{2}\sum_{{\mathbf r}} n_{{\mathbf r}}(n_{{\mathbf r}}-1).
\label{eq:Heff-2D}
\end{eqnarray}

With a four laser arrangement it is hence possible to create a density-dependent Peierls phase in the 2D lattice. 
However, since $\mathbf{\delta k}^{nm}$ must project on both $x$ and $y$ directions, there is an additional dependence of the hopping 
modulus along $y$ on $\cos(\phi n_{\mathbf r})$. 
Similar dependences of the hopping modulus appear for other choices of ${\mathbf k}_{1\dots 4}$. 

\subsection{Six-laser arrangement}
A model in which the tunneling modulus does not depend on the occupation may be attained by adding two additional lasers. This six-laser arrangement allows as well for a more flexible 
realization of density-dependent Peierls phases. We consider $\Delta_x\neq \Delta_y$, and two additional lasers $L_{5,6}$, with $\omega_5=\omega_1+\Delta_y-\Delta_x$, and 
$\omega_6=\omega_2+\Delta_y-\Delta_x$. In this way, the hopping processes (i) to (iv) along $y$ are produced, respectively, by $J_{63}$, $J_{64}$, $J_{53}$, and $J_{54}$. 
If $|\Delta_x-\Delta_y|\gg W$, the $y$ hops may be hence addressed independently  from those along $x$. 
We impose $\delta k_x^{nm}D_x=\pi$ for $n=1,2$ and $m=3,4$, and $\delta k_y^{n'm}D_y=\pi$ for $n'=5,6$ and $m=3,4$. 
A possible example is given by  ${\mathbf k}_{3,4}=-\frac{\pi}{D_x}{\mathbf e}_x$, 
${\mathbf k}_{2,5,6}=\frac{\pi}{D_y}{\mathbf e}_y$, and  
${\mathbf k}_1=\frac{\pi+\phi}{D_y}{\mathbf e}_y$. 
We choose $\Omega_2/\sqrt{3} = \Omega_1/2 = \Omega'$, $\Omega_3/\sqrt{3} = \Omega_4/2 = \Omega$ and $\Omega_6/\sqrt{3} = \Omega_5/2 = \Omega''$
and after eliminating the factors $(-1)^{j_x+j_y}$ as above, we obtain the Hamiltonian~\eqref{eq:Heff-2D_gen}
with $t_x=\frac{\Omega\Omega'}{\delta}\frac{J_x}{2\Delta_x}$ and $t_y=\frac{\Omega\Omega''}{\delta}\frac{J_y}{2\Delta_y}$ and $\phi_y=\phi r_y$.
Although also the system \eqref{eq:Heff-2D} exhibits DDSM, in the following we will for simplicity focus our analysis on Hamiltonian~\eqref{eq:Heff-2D_gen} with pure density-dependent Peierls phases.



\section{DDSM in ladders}
\label{sec:DDSMLadders}

Ladders with static fields~(i.e. Hamiltonian~\eqref{eq:Heff-2D_gen} with density-independent Peierls phases) have been recently realized in several experimental groups~\cite{Atala2014,Mancini2015,Stuhl2015} 
and studied theoretically as well~\cite{Kardar86, Orignac2001,Dhar2013, Petrescu2013,Tokuno2014,Piraud2015,Greschner2015VL}. 

An important observable in this context of density-independent fields, but also for DDSM, are currents~\cite{Atala2014}. From the continuity equation
\begin{align}
\left<\frac{dn_{{\bf r}}}{dt}\right>= {\rm i}\left<[\mathcal{H}, n_{\bf r}]\right> =-\sum_{<{\bf s}>} \mathcal{J}({\bf r}\to {\bf s})
\label{eq:contieq}
\end{align}
we can define the current $\mathcal{J}({\bf r}\to {\bf s})$ from a site ${\mathbf r}$ to a neighboring site ${\mathbf s}$.

At low fluxes the system is a MSF~\cite{Orignac2001}, characterized by the absence of rung currents. At a critical flux, 
which depends on $t_x/t_y$ and interactions, currents penetrate the rungs, form vortices, and the system becomes a VSF. Figures of typical current configurations of the VSF and MSF phases may be found in Fig.~\ref{fig:2}~(d) anticipating the discussion below. 
The MSF-VSF transition is signaled by a cusp in the chiral current $\mathcal{J}^{\rm static}_c=\mathcal{J}^{\rm static}_1-\mathcal{J}^{\rm static}_2$, with 
\begin{align}
\mathcal{J}^{\rm static}_i=\frac{\mathrm i}{L}\sum_j \langle b_{j,i}^\dag e^{-{\mathrm i}\phi_i} b_{j+1,i}-\mathrm{H.c.} \rangle
\end{align}
(in units of the hopping $t_x$) the leg currents of the ladder with density independent static magnetic Peierls phases. For bosonic systems with a finite interaction $U$, vortices may form crystals of a commensurate vortex-density $\rho_V$, which is a not-conserved quantity measuring the number of vortices per system size. Such so called vortex-lattice (VL$_{\rho_V}$) phases have been studied in weak coupling regime~\cite{Orignac2001} as well as in numerical calculations~\cite{Greschner2015VL}.\\

In the following we study the DDSM in a ladder geometry shown in Fig.~\ref{fig:1}~(a). In this situation given by the following Hamiltonian
\begin{eqnarray}
\label{eq:Hladder}
&&\!\!\!\!\!\!\!\!\!\!\!\! \mathcal{H_{L}}= -t_x\sum_{j} \left [ b_{j+1,1}^\dag e^{{\mathrm i}\phi_1 n_{j,1}} b_{j,1}+{\mathrm {H.c.}} \right ]\nonumber\\
&&\!\!\!\!\!\!\!\!\!\!\!\! -t_x\sum_{j} \left [ b_{j+1,2}^\dag e^{{\mathrm i}\phi_2 n_{j,2}} b_{j,2}+{\mathrm {H.c.}} \right ]\nonumber\\
&&\!\!\!\!\!\!\!\!\!\!\!\! - t_y  \sum_{j} \left [b_{j,2}^\dag b_{j,1} +{\mathrm {H.c.}} \right ] +\frac{U}{2} 
\sum_{i,j} n_{j,i}(n_{j,i}-1),
\end{eqnarray}

\begin{figure*}[t]
\begin{center}
\includegraphics[width=0.9\linewidth]{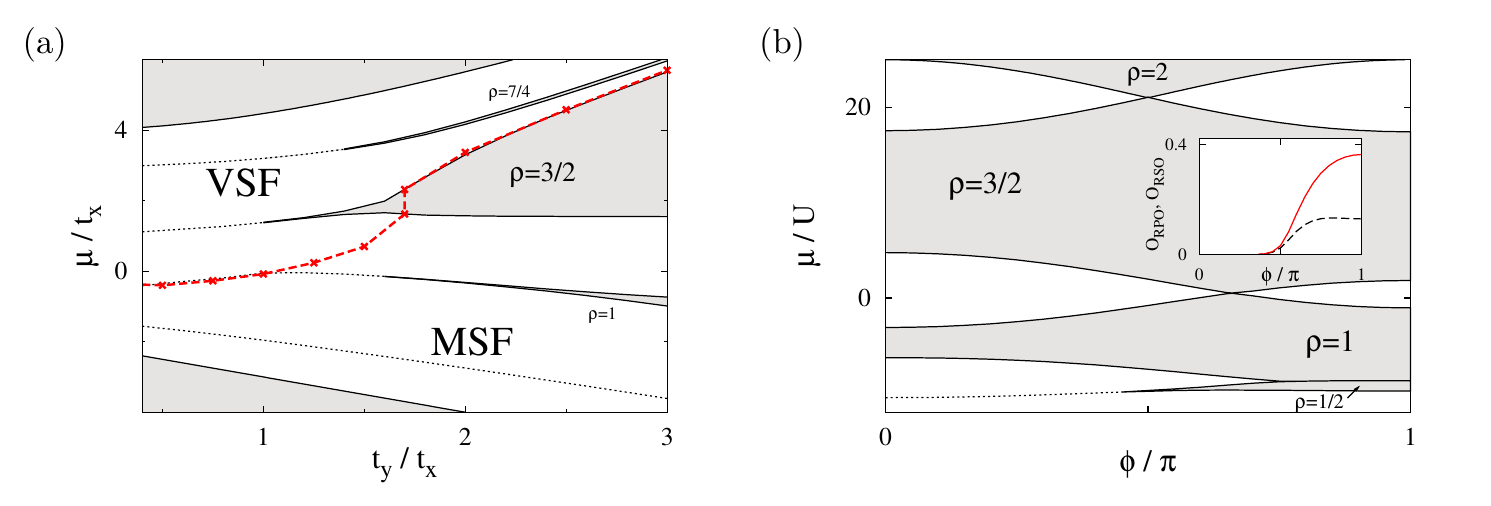}
\caption{ (a) Phase diagram for a ladder with $U=0$, $\phi_1=\pi$, and $\phi_2=0$ as function of $t_y$ and $\mu$ (in units of $t_x=1$). As discussed in the main text for large densities a VSF is realized, while for low densities the system is in a MSF phase, separated by a commensurate-to-incommensurate (with respect to the vortex density $\rho_V$) phase transition~(dashed line). Dotted lines denote lines of constant particle density $\rho=1/2, 1, 3/2$ and $7/4$, while solid lines indicate the gapped phases.
(b) Phase diagram as function of $\mu/U$ and $\phi_1=-\phi_2=\phi$ 
for $U=t_x$ and $t_y=10 t_x$ computed with DMRG. The shaded
areas denote gapped phases of $\rho=1/2, 1, 3/2$ and $2$ filling - the
white area correspond to MSF phases. As a characteristic
feature of the density-dependent fields one observes a sequence of direct transitions between
the gapped phases resulting in a macroscopic jump of density around
$\phi=\pi$, $2\pi/3$ and $\pi/2$~(see text). All gapped phases exhibit 
finite rung-string order $O_{RSO}$ and rung-parity order $O_{RPO}$ as defined in the main text. The inset shows $O_{RPO}$~(solid line) 
and $O_{RSO}$~(dashed line) for $\rho=1/2$.
}
\label{fig:largerung}
\end{center}
\end{figure*}

\subsection{MSF and VSF phases with imbalanced density}

In the limit of strong on site repulsion $U\gg J$ model~\eqref{eq:Hladder} with density dependent phases may easily be mapped onto a system of hardcore bosons without a flux for fillings $0<\rho<1$. For $1<\rho<2$ we may consider doublons $\ket{2}_{i,j}$ on top of a uniform MI-background $\prod_{i,j}\ket{1}_{i,j}$ as hardcore particles, which, however, now experience a finite flux $\phi=\phi_1-\phi_2$, such that the effective Hamiltonian in this limit may be written as
\begin{eqnarray}
\label{eq:Hhardcore}
&&\!\!\!\!\!\!\!\!\!\!\!\! H_{\rm hardcore}^{\rho>1}= -2 t_x\sum_{i,j} \left [ c_{j+1,i}^\dag e^{{\mathrm i}\phi i} c_{j,i}+{\mathrm {H.c.}} \right ]\nonumber\\
&&\!\!\!\!\!\!\!\!\!\!\!\! - 2 t_y  \sum_{j} \left [c_{j,2}^\dag c_{j,1} +{\mathrm {H.c.}} \right ],
\end{eqnarray}
with $c_{i,j}$ ( $c_{i,j}^\dagger$) being the creation (annihilation) operator of a doublon on site $(i,j)$. 
Thus in the strongly interacting regime $U\gg J$, model~\eqref{eq:Hladder} is expected to reproduce the physics of hardcore bosons in a magnetic static field exhibiting MSF and VSF phases as discussed in detail in~\cite{Piraud2015}.

A qualitative insight on the physics induced by the occupation-dependent Peierls phases away from that limit is provided by a simple mean-field 
decoupling of the tunneling terms (between neighboring sites ${\bf r}$ and ${\bf r}'$) in Eq.~\eqref{eq:Hladder}:
Since $(b_{{\bf r}}^\dag)^3=0$, then 
$b_{{\mathbf r}'}^\dag e^{{\mathrm i}\phi n_{{\mathbf r}}} b_{{\mathbf r}}=
b_{{\mathbf r}'}^\dag (1+(e^{{\mathrm i}\phi}-1)n_{{\mathbf r}})  b_{{\mathbf r}}$. Using the decoupling
$b_{{\mathbf r}'}^\dag n_{{\mathbf r}} b_{{\mathbf r}}\simeq 2\kappa({\mathbf r}',{\mathbf r}) (n_{{\mathbf r}}-{\bar n}_{\mathbf r})+2{\bar n}_{{\mathbf r}}b_{{\mathbf r}'}^\dag b_{{\mathbf r}}$ we obtain
\begin{eqnarray}
\label{eq:MF}
\!\!\!\!\!\!\!\! b_{{\mathbf r}'}^\dag e^{{\mathrm i}\phi n_{{\mathbf r}}} b_{{\mathbf r}} + {\mathrm {H.c.}}
\!\! &\simeq&\!\!   [(1+2{\bar n}_{{\mathbf r}}(e^{{\mathrm i}\phi}-1) b_{{\mathbf r}'}^\dag b_{{\mathbf r}} +{\mathrm {H.c.}}]  \\
\!\! &+& \!\! [2\kappa({\mathbf r}',{\mathbf r})(e^{{\mathrm i}\phi}-1) +{\mathrm {c.c.}}] (n_{{\mathbf r}}-{\bar n}_{\mathbf r}), \nonumber
\end{eqnarray}
with $\bar n_{\mathbf r}\equiv \langle n_{{\mathbf r}} \rangle$, and $\kappa({\mathbf r}',{\mathbf r})\equiv \langle b_{{\mathbf r}'}^\dag b_{{\mathbf r}} \rangle$.
The first term at the right hand side~(rhs) of Eq.~\eqref{eq:MF} results in an effective Peierls phase. 
A density-dependent effective flux is hence given by the phase accumulated when encircling a plaquette. For model~\eqref{eq:Hladder} the effective flux 
is uniform for homogeneous $\bar n_{i,j}=\bar n_{i}$. The second term at the rhs introduces a shift of the local chemical 
potential, which is leg-dependent for  $\phi_2\neq -\phi_1$ in Eq.~\eqref{eq:Hladder}.
We hence expect from this simple argument that the occupation-dependent Peierls phase introduces an interplay between density-dependent fields and density imbalance between the legs.
We employ below density matrix renormalization group (DMRG)~\cite{Schollwoeck11} calculations to confirm this insight. In these calculations we use system sizes up to $100$ rungs with open boundary conditions keeping up to $1000$ matrix states.

As in Ref.~\cite{Atala2014} we monitor the chiral current, $\mathcal{J}_c=\mathcal{J}_1-\mathcal{J}_2$, where the leg currents 
are now defined as 
\begin{align}
\mathcal{J}_i=\frac{\mathrm i}{L}\sum_j \langle b_{j,i}^\dag e^{-{\mathrm i}\phi_i n_{j,i}} b_{j+1,i}-\mathrm{H.c.} \rangle
\end{align}
in units of $t_x/\hbar$, with $N_b$ the number of bonds along the leg. 
The density-dependent effective flux induces a characteristic dependence of $\mathcal{J}_c$ on chemical potential, $\mu$, as shown in Figs.~\ref{fig:2}~(b) and (c)  
for $\phi_1=0.8\pi$, and $\phi_2=0$~(similar results occur for other parameter values). 
In addition to MI phases, we observe two different SF regimes. For a given $t_x/U$ there is a critical $\mu$, at which $J_c$ presents a cusp~(see Fig.~\ref{fig:2}(c)), indicating a MSF-VSF transition induced by the increasing effective flux for growing lattice filling $\rho$. 

This transition is as well characterized by a kink in the equation of state $\rho(\mu)$ which signals a change in the number of gapless modes of the system: The MSF phase has a gap in the antisymmetric (or naively ``Vortex''-) sector~\cite{Orignac2001}, while the symmetric (``charge'') sector remains gapless. The number of gapless modes is also reflected by the central charge $c$, which is $c=1$ in the MSF phase and $c = 2$ in the VSF phase. We verify this by numerically extracting $c$ from the scaling of the entanglement entropy~\cite{Piraud2015}.

Figures~\ref{fig:2}(a) and~(c) also show that the occupation-dependent Peierls phase leads to a marked density imbalance, 
$\Delta n= 2(\bar n_2-\bar n_1)/(\bar n_2+\bar n_1)$. Three important points should be noted. First, although $|\Delta n|$ is particularly large in the MSF, it is non-vanishing as well within the VSF. Second, $\Delta n$ presents a kink at the MSF-VSF transition. Third, although $\Delta n$ results from the explicitly broken symmetry between the legs in 
Eq.~\eqref{eq:Hladder}, its sign depends non-trivially on $\mu$ or $\rho$. Figure~\ref{fig:2}(c) shows that $\Delta n$ may change 
its sign going through a balanced point, $\Delta n=0$.



\subsection{Strong rung-coupling limit} 

As for the case of static magnetic fields where the MSF-VSF transition has been explored for a fixed flux in Ref.~\cite{Atala2014} as function of $t_y/t_x$, also for the DDSM the rung hopping strength constitutes an important degree of freedom. In Fig.~\ref{fig:largerung}~(a) we study the phase diagram, in particular the commensurate-to-incommensurate MSF-VSF transition, on $t_y$ for $U=0$, $\phi_1=\pi$, and $\phi_2=0$. Interestingly we basically observe two different regimes: For small interchain couplings $t_y/t_x\lesssim 1$ the MSF-VSF boundary (dashed line) is located close to unit filling, however, as $t_y/t_x\gtrsim 1$ it shifts quickly to larger densities  $\rho \sim 3/2$.

For $\phi_1-\phi_2\neq\pi$ above a critical value of $t_y/t_x$ the VSF phase may vanish. 
In the strong-rung coupling limit $t_y/t_x\gg 1$ several gapped band insulating phases at commensurate fillings $\rho=1$ and $3/2$ are stabilized. The extent of the MI-phases strongly increases with $t_y/t_x$ as discussed below. For the parameters of Fig.~\ref{fig:largerung}~(a) a MI-phase at $\rho=1/2$ is suppressed. Additionally one may observe a gapped charge density wave phase at filling $\rho=7/4$ (see \cite{Piraud2015} for a detailed discussion of similar phases at $1/4$ filling for density-independent synthetic magnetism).
Apart from the SF-phases also the MI-phases may be of Meissner-MI (in Fig.~\ref{fig:2}(a) for $\rho=1/2$ and $\rho=1$) and of vortex-MI (for $\rho=3/2$) types as discussed in 
Ref.~\cite{Piraud2015}. Both exhibit a mass gap, however, the vortex-MI still has gapless mode (i.e. the neutral gap in the manifold of constant particle number vanishes) while the Meissner-MI phase is completely gapped.\\

For $t_y\gg U, t_x$ the ladder reduces to an effective rung-chain model with intriguing novel features due to the 
density-dependent Peierls phases. 
We may then map to rung-states $|\tilde{N}\rangle$ with a fixed occupation, of ${\tilde N}=0,1,2,3,4$ particles, on each rung.
For the particular case of $\phi_1=-\phi_2=\phi$ at $t_x=0$, the ground-states of the decoupled rungs are the rung states 
\begin{eqnarray}
&&|\tilde 0\rangle\equiv {0\choose 0}, ~ \mu<-t_y, \notag\\
&&|\tilde 1\rangle\equiv \frac{1}{\sqrt{2}}\left [{1\choose 0}+{0\choose 1} \right ], ~ -t_y<\mu<-t_y+\frac{U}{2},\notag\\ 
&&|\tilde 2\rangle\equiv \frac{1}{2}\left [{2\choose 0}+\sqrt{2}{1\choose 1}+{0\choose 2} \right ], ~ -t_y+\frac{U}{2}<\mu<\frac{U}{2}, \notag
\end{eqnarray}
where the notation $n_1 \choose n_2$, denotes the rung state
with $n_1$~($n_2$) particles in the upper~(lower) leg.
At low $\rho$ (in the vicinity of $\mu\sim -t_y$), for which ${\tilde N}>2$ are irrelevant, the effective rung-chain model becomes of the form:
\begin{eqnarray}
\mathcal{H}&=&-t_x\sum_j \left [ B_j^\dag (1-\sin^2(\phi/2) N_j) B_{j+1} + \mathrm {H.c.} \right ]  \nonumber \\
 &+& \frac{U}{4}\sum_j N_j(N_j-1)-(\mu+t_y)\sum_j N_j,
 \label{eq:HRung}
\end{eqnarray}
where $B_j$ are bosonic operators in the space $\{ |\tilde{0}\rangle, |\tilde{1}\rangle, |\tilde{2}\rangle \}$, and $N_j=B_j^\dag B_j$. 
Note that inter-rung hops $|\tilde{0}\rangle |\tilde{1}\rangle \leftrightarrow |\tilde{1}\rangle |\tilde{0}\rangle$ and $|\tilde{1}\rangle |\tilde{1}\rangle \leftrightarrow |\tilde{0}\rangle |\tilde{2}\rangle$ have an amplitude $t_x$, 
whereas $|\tilde{1}\rangle |\tilde{1}\rangle \leftrightarrow |\tilde{2}\rangle |\tilde{0}\rangle$ and $|\tilde{2}\rangle |\tilde{1}\rangle \leftrightarrow |\tilde{1}\rangle |\tilde{2}\rangle$ have an amplitude $t_x\cos^2(\phi/2)$. 
The latter rate vanishes for $\phi=\pi$. 
As a result a direct transition occurs for finite $t_x$ between the gapped 
phases at fillings $\rho=1/2$ and $1$, i.e. $\tilde N=1$ and $2$, with an infinite compressibility and a macroscopic density jump~(Fig.~\ref{fig:largerung}(b)). 
Similarly, direct transitions occur between gapped phases with $\rho=1$ and $3/2$ (at $\phi=2\pi/3$), and $3/2$ and $2$~(at $\phi=\pi/2$).

The presence of the density-dependent phases results in a broken 
space-inversion symmetry, since the amplitudes of $|\tilde{1}\rangle |\tilde{1}\rangle \leftrightarrow |\tilde{2}\rangle |\tilde{0}\rangle$ and $|\tilde{1}\rangle |\tilde{1}\rangle \leftrightarrow  |\tilde{0}\rangle |\tilde{2}\rangle$ are not equal. As discussed in Ref.~\cite{Greschner2014a} the broken space inversion symmetry may result in the exotic situation of simultaneous presence of both nonlocal parity- and string-order in the insulating MI-phases. The MI phase of a usual 1D Bose Hubbard model is characterized by a finite hidden parity order due to bound particle hole pairs that has been observed in experiments with single site resolution~\cite{Endres2011}. A non-vanishing 
string order, but vanishing parity order, characterizes the Haldane insulator, predicted in polar lattice gases~\cite{DallaTorre2006,Berg2008} and bosons in frustrated lattices~\cite{Greschner2013b}.
The explicit expressions in the effective rung-state model may be borrowed from the corresponding orders of a spin $S=1$ chain~\cite{Berg2008}: We define the rung-parity-order 
 $O_{RPO} \equiv \lim_{|i-j|\to\infty}  \langle (-1)^{\sum_{i<k<j}\delta N_k} \rangle$ ~(with $\delta N_k=\tilde N -N_k$) and the rung-string-order
$O_{RSO} \equiv \lim_{|i-j|\to\infty}  \langle \delta N_i (-1)^{\sum_{i<k<j} \delta N_k} \delta N_j \rangle$. 
For $\phi=0$ the Mott phases of the rung-chain model~\eqref{eq:HRung} present finite rung-parity-order $O_{RPO}$ but vanishing rung-string-order $O_{RSO}$. 
Due to the density depended phases the Mott rung phases acquire a simultaneous finite $O_{RPO}$ and $O_{RSO}$, as may be seen in the inset of Fig.~\ref{fig:largerung}~(b).

\begin{figure}[t]
\begin{center}
\includegraphics[width=1.\columnwidth]{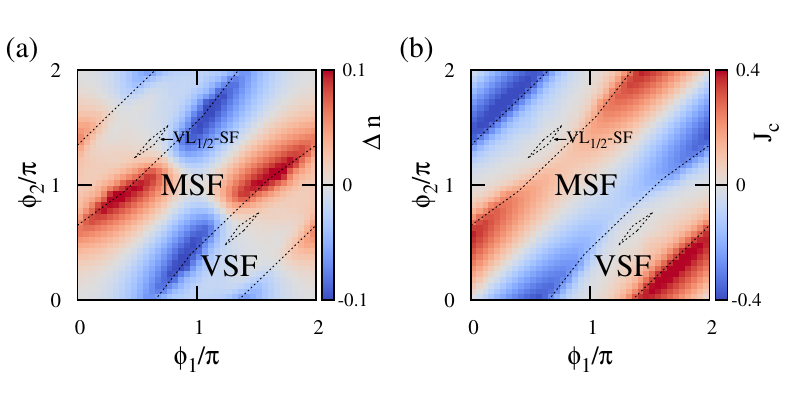}
\caption{ (a) Density imbalance $\Delta n$ and (b) chiral current $\mathcal{J}_c$  function of $\phi_1$ and $\phi_2$ for $t_x=t_y$, $U=t_x$ and $\rho=1.25$ as obtained by DMRG calculations. In addition to the MSF and VSF phases a small vortex lattice phase at vortex-density $\rho_V=1/2$, VL$_{1/2}$-SF may be observed. Dashed lines indicate the  phase boundaries from the VSF to the MSF and VL$_{1/2}$-SF phases.}
\label{fig:phi1ph2}
\end{center}
\end{figure}
\subsection{Symmetries and vortex-lattice phases}

Density independent magnetic fields are up to a gauge transformation completely  defined by the net flux per unit-cell of the lattice. Due to its operator-nature this is not true for the case the density dependent Peierls phases. Indeed as may be seen in Fig.~\ref{fig:phi1ph2} the phase diagram may significantly depend on the values of both phases $\phi_1$ and $\phi_2$ of model~\eqref{eq:Hladder}. While the MSF-VSF phase boundary mainly just depends on the total effective flux $\phi_1-\phi_2$, only in the vicinity of $\phi_1\simeq -\phi_2\simeq \pi/2$, where also density imbalance $\Delta n$ vanishes, we observe a vortex-lattice phase at vortex-density $\rho_V=1/2$~(VL$_{1/2}$ phase). Apart from the characteristic staggered pattern of the currents as shown for the case of static magnetic fields e.g. in \cite{Greschner2015VL}, the VL$_{1/2}$ phase may be discriminated from the VSF phases by the calculation of the central-charge, which is $c=1$ in vortex-lattice phases. For strong phase-imbalances $\phi_1=0,\phi_2=\phi$ as in Fig.~\ref{fig:2} no vortex-lattice phases are observed.

Note that the phase diagram is symmetric with respect to inversion of the phases $I_1: (\phi_1, \phi_2) \to (-\phi_1, -\phi_2)$ and exchange of the two legs of the ladder $I_2: (\phi_1, \phi_2) \to (\phi_2, \phi_1)$ for $\phi_1, \phi_2 \in [0,2 \pi)$. The density imbalance $\Delta n$ (Fig.~\ref{fig:phi1ph2}~(a)) is (anti)symmetric with respect to $I_1$($I_2$). The chiral current(Fig.~\ref{fig:phi1ph2}~(b)) is an antisymmetric quantity under both $I_1$ and $I_2$.

\section{Two-dimensional square lattices}
\label{sec:2D}

%
\begin{figure*}[t]
\begin{center}
\includegraphics[width=1.\linewidth]{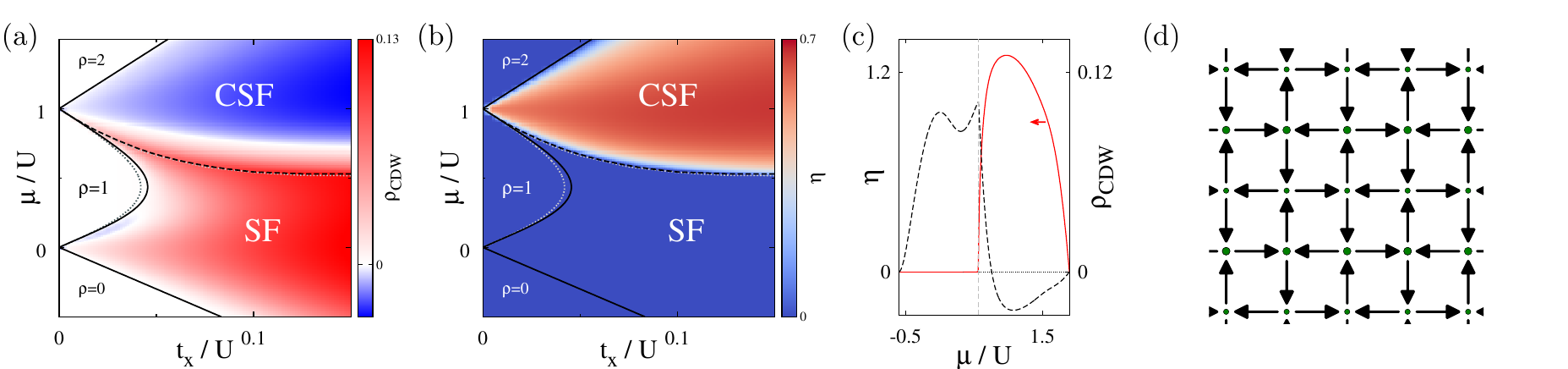}
 \caption{~(a) and (b): Ground state phase diagram of model (\ref{eq:Heff-2D_gen}) in 2D with $\pi$-phases $\phi_{\mathbf{r}}=j\pi$ and $t_x=2t_y$ computed with CBMFT with clusters of size $L_x\times L_y=2\times 2$ (black lines) and $4\times 2$ (dotted grey lines). Solid lines mark the boundaries of the MI, while the the onset of a finite chiral bond order parameter, $\eta$, signaling the CSF-SF transition, is marked with dashed lines and the dashed curve marks. The color code indicates (a) the charge density wave order parameter $\rho_{CDW}$ and (b) the chiral bond order parameter $\eta$, as defined in the main text. Note that the CBMFT results remain stable under increasing of the cluster size. (c) Chiral bond order parameter $\eta$~(solid line) and charge density wave order parameter $\rho_{CDW}$~(dashed line) for a cut in the phase diagram along $t_x/U=0.1$. (d) Typical current and density configuration of the CSF phase. The size of the circles is proportional to the onsite-density, the lengths and widths of the arrows encode the strength of the local currents.
\label{fig:2d} }
\end{center}
\end{figure*}

We now extend our study to the effect of DDSM to two dimensional square lattices. Equivalently to the ladder case, we show that the occupation-dependent Peierls phase induces a non trivial interplay between the density-dependent phases and density modulations. As a first approach, we focus on the limiting case of $\pi$-phases, i.e. $\phi_{\mathbf{r}}=j \pi$, for which the Peierls phase in Hamiltonian~\eqref{eq:Heff-2D_gen} takes the simpler form,
\begin{equation}
e^{{\rm i}\phi_{\mathbf{r}} n_{\mathbf{r}}} = (-1)^{j n_{\mathbf{r}}}
\end{equation}
Despite this simplification, the Peierls phase of the hopping along the $x$-direction still rends the 2D model (\ref{eq:Heff-2D_gen}) to be intrinsically frustrated and thus highly non-trivial to approach from a computational perspective. In the following, we will use the composite boson mean-field theory (CBMFT)~\cite{Huerga2013,Huerga2014}, which is a useful tool to unveil strongly correlated phases of spin and boson lattice models where other methods face significant problems.

CBMFT is based on the use of clusters of the original degrees of freedom as the basic degrees of freedom that contain the necessary quantum correlations to describe the phases emerging in the system under study. In practice, we tile the lattice into clusters of equal size, in such a way that each site $\mathbf{r}$ of the original 2D lattice belongs to a unique cluster. The tiling is performed preserving most of the symmetries of the model. Each quantum state of each cluster can be represented by the action of a creation {\it composite boson} (CB) over a CB vacuum. Being the mapping relating the original bosons $\{b_{\mathbf{r}}^{\dag},b_{\mathbf{r}}\}$ to the new CBs canonical \cite{Huerga2013}, one can rewrite (\ref{eq:Heff-2D_gen}) in terms of CBs and approach it by standard many-body techniques, with the advantage that short-range quantum correlations are exactly computed by construction. 

Here we will use the CB Gutzwiller ansatz, a simplest product of uncorrelated cluster wave functions,
\begin{equation}
\ket{\Phi}= \prod_{\mathbf{R}} a^{\dag}_{\mathbf{R},{\sf g}} \ket{0_{CB}}=\prod_{\mathbf{R}}\ket{{\sf g}}_{\mathbf{R}}\label{eq:cb_gutz}
\end{equation}
where $a_{\mathbf{R},{\sf g}}^{\dag}$ is the creation CB associated to the cluster $\mathbf{R}$ in the state $\ket{{\sf g}}_{\mathbf{R}}=\sum_{\mathbf{n}} U_{\mathbf{n}}^{(\mathbf{R})} \ket{\mathbf{n}}_{\mathbf{R}}$, where $\mathbf{n}$ refers to a cluster configuration in the occupation basis. The amplitudes $U_{\mathbf{n}}^{(\mathbf{R})}$ are then determined upon variational minimization of the energy. In the homogeneus case, i.e. $U_{\mathbf{n}}^{(\mathbf{R})}=U_{\mathbf{n}}$, this variational determination is equivalent to exactly diagonalize a unique cluster with open boundary conditions and a set of self-consistently defined mean-fields acting on its borders \cite{Huerga2014}. 

The CB Gutzwiller ansatz (\ref{eq:cb_gutz}) allows to compute observables and order parameters in a systematic way. In particular, the energy obtained is variational, and the ground state phase diagram can be obtained by monitoring the ground state energy and its derivatives. In addition, low lying excitations over the ground state can be analyzed within the CBMFT framework self-consistently \cite{Huerga2013}. Nevertheless, this analysis is out of the scope of the present work.

We define a $(0,\pi)$ charge density wave (CDW) order parameter, 
$\rho_{CDW}=\sum_{\mathbf{r}} e^{-{\mathrm i} \pi j} \bra{\Phi} n_{\mathbf{r}} \ket{\Phi}/N
$, which computed with an homogeneus CB Gutzwiller ansatz $\ket{\sf g}$ takes the form,
\begin{equation}
\rho_{CDW}=\frac{1}{L_x L_y}\sum_{\mathbf{r}\in \square} e^{-{\mathrm i} \pi j} \bra{\sf g} n_{\mathbf{r}} \ket{\sf g}.
\end{equation}

and the bond-chiral order parameter,
\begin{equation}
\eta= \frac{1}{N_b}\sum_{\langle {\bf r},{\bf r'} \rangle}|\langle\Phi |\mathcal{J}_{{\bf r},{\bf r'}}|\Phi\rangle|
\end{equation}
where $N_b$ is the number of bonds, and the currents $\mathcal{J}_{\mathbf{r,r'}}$ are defined through the continuity equation~\eqref{eq:contieq}.

Figures~\ref{fig:2d}~(a) and (b) show CBMFT results of the ground-state phase diagram of Eq.~(\ref{eq:Heff-2D_gen}) with $\pi$-phases and clusters of size $L_x\times L_y=2\times2,~4\times 2$. These sizes preserve the periodicity imposed by the Peierls phase with effective $\pi$-flux. In order to enhance the non-trivial hopping of bosons along the $x$-direction we have set $t_x=2t_y$. 

As we can see in Figs.~\ref{fig:2d}~(a) and (b), the system presents the usual MI lobes of integer density for small values of the hopping $t_x$. For bigger values of the hopping, the ground state presents superfluid order, characterized by a nonvanishing condensate density, $\rho_0=\bra{\Phi}b^{\dag}_{\mathbf{k=0}}b_{\mathbf{k=0}}\ket{\Phi}/N$ (not shown). In particular, a SF phase with modulated density and vanishing bond-chiral order emerges for $\rho<1$, while for $\rho>1$ the ground state is a CSF characterized by having nonvanishing bond currents in a pattern of fully stacked checkerboard pattern of vortices and anti-vortices~(Fig.~\ref{fig:2d}~(d)). Notice that the CSF phase is nothing but a limiting case of the VSF previously described in the ladder case, when the vortices are of the size of a single plaquette in the square lattice. In addition, the CSF has nonvanishing density modulations. 
Were the Peierls phase density-independent, all the superfluid region would have nonvanishing chiral order, and the density modulations would disappear (not shown). Thus, the density-dependence in the Peierls phase has the effect of inducing finite density modulations and reducing the region with nonvanishing chiral order to that of $\rho>1$.

The phase transitions are in all cases found to be of second order, signaled by discontinuities in the second order derivative of the energy with respect to the chemical potential. The continuous vanishing of the BEC order parameter (SF-MI transition) and the bond-chiral order $\eta$ (CSF-SF transition) also supports this assumption. Moreover, the phase diagram remains stable under increasing of the cluster size, as the CBMFT-4$\times$2 includes minor quantitative corrections to the phase borders of the CBMFT-2$\times$2. In particular, the CSF-SF phase boundary obtained with 2$\times$2 and 2$\times$4 basically overlap~(Figs.~\ref{fig:2d}~(a) and (b)). 

Comparing the phase diagrams for the ladder~(Fig. \ref{fig:2}) and the 2D square lattice~(Fig. \ref{fig:2d}), we observe that the modulated SF can be considered as the bulk counterpart of the MSF appearing in the ladder geometry.



\section{Dynamically probing the density-dependent field}
\label{sec:Dynamics}

DDSM results in an intriguing dynamics that may be easily probed experimentally. 
We illustrate this point with the particular case of the ladder model~\eqref{eq:Hladder} with $\phi_1=-\phi_2=\phi$ and $t_x=t_y$. 
We are interested in the dynamics of a defect (formed by either a doubly-occupied site, i.e. a doublon, or an empty site, i.e. a holon) created in a MI with $\rho=1$, initially at site $(1,j=0)$. Note that this initial condition is chosen for simplicity of the analysis. The initial doublon or holon may be created in a more delocalized region of the ladder. The relevant conclusions about the expansion dynamics would be unaffected. Similar dynamics has been studied recently in the context of Bose Hubbard models without gauge fields \cite{Andraschko2015} and may be observed in experiments with single site resolution~\cite{Bakr2010, Wuertz2009, Cheneau2012}.

\begin{figure}[b]
\centering
\includegraphics[width=1.\columnwidth]{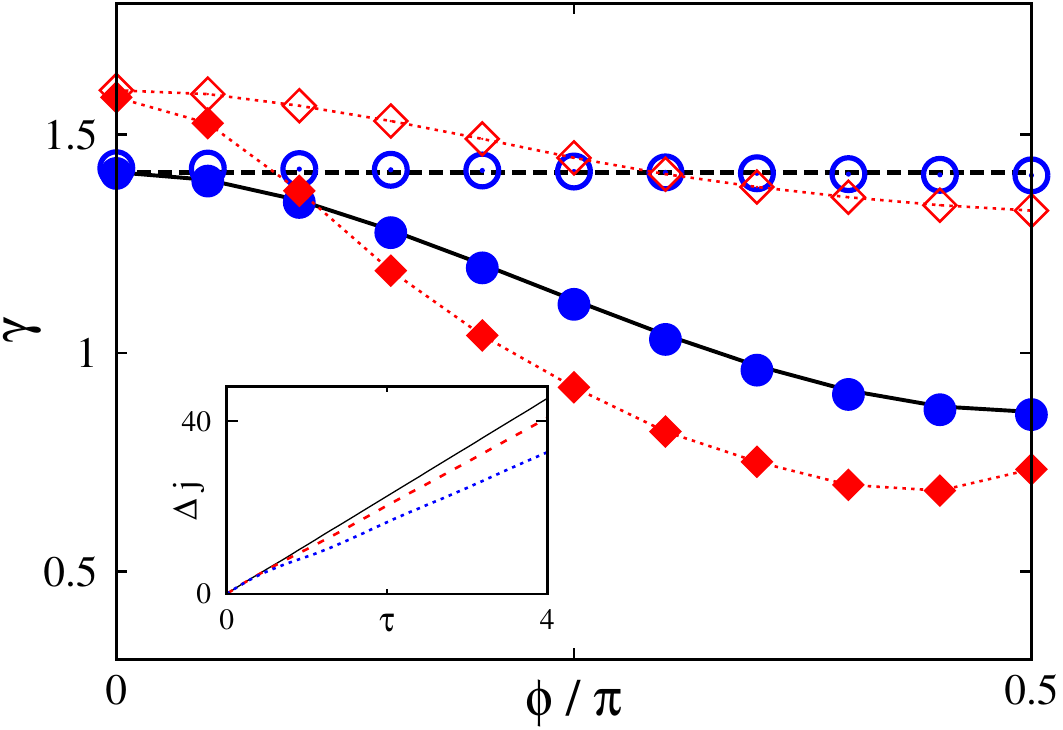}
\caption{
 Expansion coefficient $\gamma$ of a defect along the ladder as a function of the effective flux $\phi$ for $t_x=t_y$, and $U/t_x=50$~(circles) and $U/t_x=10$~(diamonds). Hollow~(filled) symbols denote the t-DMRG results for the holon~(doublon) expansion. Dashed~(solid) curves denote single-particle (exact diagonalization) results for holons~(doublons), which 
match well with the t-DMRG results for large $U/t_x$. The inset depicts typical linear expansions of $\Delta j(\tau)$ for a doublon at $U\to\infty$ and $\phi/\pi=0$ (solid line), $0.5$ (dashed line) and $1$ (dotted line).
}
\label{fig:4}
\end{figure}

For $U\gg t_x$ quantum (particle/hole) fluctuations of the MI
are irrelevant, and the defect expansion is like that of a single particle with a hopping $t_x$~($2 t_x e^{\pm{\mathrm i} \phi}$) for the holon~(doublon). 
Both holon and doublon  propagate ballistically along the ladder, i.e. $\Delta j(\tau) = \sqrt{\langle j^2 \rangle (\tau)} \sim \gamma \tau$~(we consider below the time $\tau$ in units 
of $\hbar/t_x$ for holons and $\hbar/2t_x$ for doublons). The expansion 
coefficient $\gamma$ is however markedly different. Holons do not experience any magnetic flux, and thus they propagate with a $\phi$-independent 
$\gamma=\sqrt{2}$. In contrast, doublons experience a flux $2\phi$ and their trajectories are partially diverted by cyclotron motion. Hence 
$\gamma$ decreases with $\phi$~(Fig.~\ref{fig:4}). The inset of Fig.~\ref{fig:4} depicts examples of $\Delta j(\tau)$ for different $\phi$.
This situation has to be contrasted with the case of density independent magnetic fields. Here holons and doublons will both experience the same magnetic flux $2\phi$  and propagate - up to a factor $2$ due to bosonic enhancement - in the same way.

For lower $U/t$ quantum fluctuations become relevant altering the defect expansion in an intriguing way.
A perturbative treatment of the role of particle-hole fluctuations offers an instructive starting point of the study. Up to second-order one virtual doublon-holon pair may be created and annihilated which mediate new hoppings of the initial holon~(doublon) of the form:
\begin{eqnarray}
\mathcal{H}^{(2)}&=&\frac{-2t_x^2}{U}\sum_{i,j} \left [ \alpha_i |j+2,i\rangle\langle j,i| +\mathrm {H.c.}\right ] \nonumber \\
&-&\frac{2t_xt_y}{U} \sum_{i,j}\left [  \beta_i |j+1,k\neq i\rangle\langle j,i| +\mathrm {H.c.}\right ],
\label{eq:Hextra}
\end{eqnarray}
where $|i,j\rangle$ denotes a defect at site $(i,j)$, $\alpha_i\equiv e^{{\mathrm i}\phi_i}$~($e^{-{\mathrm i}\phi_i}$), $\beta_i=1+e^{{\mathrm i}\phi_i}$~($1+e^{-{\mathrm i}\phi_{k\neq i}}$) for doublons~(holons). In order to study the influence of quantum fluctuations beyond perturbation theory we perform t-DMRG calculations~\cite{Schollwoeck11}, with system sizes up to $100$ rungs keeping up to $1000$ matrix states.

As shown in Fig.~\ref{fig:4} for $U/t_x=10$, for $\phi=0$ fluctuations speed up defect expansion; the expansion coefficient may reach values $\gamma\simeq 1.6$.  
This is intuitively clear since there are more processes expanding the defect along the ladder. This remains true for small $\phi$. However, the peculiar phase 
dependence of the extra terms~\eqref{eq:Hextra}, modifies as well the effective magnetic flux experienced by the doublons. 
Indeed, for a sufficient large $\phi$, fluctuations slow down the doublon expansion, i.e. they strengthen the cyclotron motion diverting the doublon expansion, 
corresponding to an increase of  the effective 
magnetic field experienced by the doublons. Moreover, quantum fluctuations make holon expansion 
$\phi$--dependent due to virtual doublons. For sufficiently large $\phi$, fluctuations slow down the holon expansion, i.e. holons experience 
an effective cyclotron motion induced by quantum fluctuations of the MI substrate.



\section{Summary}
\label{sec:Summary}

Raman-assisted hopping may be used to induce density-dependent synthetic magnetism in cold lattice gases. In one dimensional systems this results in the interesting possibility of studying the anyon model~\cite{Greschner2015AHM}. For ladders and 2D square lattices we have shown that these fields lead to a rich ground-state physics characterized by the non-trivial interplay between density modulations and chirality. In two-leg ladders it is characterized by a density-driven Meissner- to vortex-superfluid transition. 
Moreover, DDSM significantly affects the dynamics of particles in the lattice, leading in particular to an intriguing expansion dynamics for 
doublons and holons in a MI, which presents a remarkable dependence on quantum fluctuations and may be used to reveal experimentally the DDSM. 

Although we have focused on ladders and 2D square lattices, similar ideas may be applied to 
more general lattices, opening interesting possibilities for the realization of density-induced geometric frustration.
In this work we discussed bosonic particles in the presence of DDSM. In Ref.~\cite{Greschner2015AHM} it is shown that also fermionic species may be a useful candidate for the realization of DDSM in cold atom experiments, since here the spurious (vi)-(ix) processes of section II E identically vanish. While in one dimensional systems this can be exploited to study the anyon Hubbard model, in two and higher dimension a significantly different model is realized.
These possibilities will be examined in forthcoming works.\\

\begin{acknowledgments}
We thank J. Dukelsky, G. Ortiz, 
M. di Liberto, C. de Morais-Smith, T, Mishra, T. Vekua, A. Eckardt, M. Dalmonte, M. Roncaglia, P. \"Ohberg, M. Valiente, M. Lewenstein, T. Grass, 
and B. Juli\'a-D\'\i az for enlightening discussions. 
We acknowledge support by the cluster of excellence QUEST, the DFG Research Training Group 1729, the 
SUTD start-up grant (SRG-EPD-2012-045) and the Spanish Ministry of Economy and
Competitiveness through grants FIS-2012-34479, BES-2010-031607 and
EEBB-14-09077. 
\end{acknowledgments}

\bibliographystyle{prsty}

\begin{thebibliography}{}

\bibitem{VonKlitzing1980}  K. von Klitzing, G. Dorda, and M. Pepper, Phys. Rev. Lett. {\bf 45}, 494 (1980).
\bibitem{Tsui1982} D. C. Tsui, H. L. Stormer, and A. C. Gossard, Phys. Rev. Lett. {\bf 48}, 1559 (1982).
\bibitem{Laughlin1983} R. B. Laughlin, Phys. Rev. Lett. {\bf 50}, 1395 (1983).
\bibitem{Wilczek1982} F. Wilczek, Phys. Rev. Lett. {\bf 49}, 957 (1982).
\bibitem{Kane2005} C. L. Kane and E. J. Mele, Phys. Rev. Lett. {\bf 95}, 226801 (2005).
\bibitem{Bernevig2006} B. A. Bernevig and S.-C. Zhang, Phys. Rev. Lett. {\bf 96}, 106802 (2006).
\bibitem{Bloch2008} I. Bloch, J. Dalibard, and W. Zwerger, Rev. Mod. Phys. {\bf 80}, 885 (2008).
\bibitem{Dalibard2011} J. Dalibard, F. Gerbier, G. Juzeliunas, and P. \"Ohberg, Rev. Mod. Phys. {\bf 83}, 1523 (2011).
\bibitem{Goldman2014} N. Goldman, G. Juzeliunas, P. \"Ohberg, and I. B. Spielman, Rep. Prog. Phys. {\bf 77}, 126401 (2014).
\bibitem{Lin2011}  Y. J. Lin {\it et al.},  Nature Physics {\bf 7}, 531 (2011).
\bibitem{Lin2011b}  Y. J. Lin, K. Jim\'enez-Garc\'ia and I. B. Spielman,  Nature Physics {\bf 471}, 83 (2011).
\bibitem{Aidelsburger2013} M. Aidelsburger, M. Atala, M. Lohse, J. T. Barreiro, B. Paredes, and I. Bloch, Phys. Rev. Lett. {\bf 111}, 185301 (2013).
\bibitem{Miyake2013} H. Miyake, G.A. Siviloglou, C.J. Kennedy, W.C. Burton and W. Ketterle, Phys. Rev. Lett. {\bf 111}, 185302 (2013).

\bibitem{Orignac2001} E. Orignac and T. Giamarchi, Phys. Rev. B \textbf{64}, 144515 (2001).
\bibitem{Atala2014} M. Atala, M. Aidelsburger, M. Lohse, J. T. Barreiro, B. Paredes, I. Bloch, Nature Physics {\bf 10}, 588 (2014).
\bibitem{Mancini2015} M. Mancini {\it et al.},  arXiv:1502.02495.
\bibitem{Stuhl2015} B. K. Stuhl {\it et al.}, arXiv:1502.02496v1.

\bibitem{Levin2005} M. Levin and X. G. Wen, Rev. Mod. Phys. {\bf 77}, 871 (2005).
\bibitem{Kogut1983}  J. Kogut, Rev. Mod. Phys. {\bf 55}, 775 (1983).

\bibitem{Cirac2010} J. I. Cirac, P. Maraner, and J. K. Pachos, Phys. Rev. Lett. {\bf 105}, 190403 (2010).
\bibitem{Zohar2011} E. Zohar and B. Reznik, Phys. Rev. Lett. {\bf 107}, 275301 (2011).
\bibitem{Kapit2011} E. Kapit and E. Mueller, Phys. Rev. A {\bf 83}, 033625 (2011).
\bibitem{Zohar2012} E. Zohar, J. I. Cirac, and B. Reznik, Phys. Rev. Lett. {\bf 109}, 125302 (2012).
\bibitem{Banerjee2012} D. Banerjee, M. Dalmonte, M. M\"uller, E. Rico, P. Stebler, U.-J. Wiese, and P. Zoller, Phys. Rev. Lett. {\bf 109}, 175302 (2012).
\bibitem{Zohar2013} E. Zohar, J. I. Cirac, and B. Reznik, Phys. Rev. Lett. {\bf 110}, 055302 (2013).
\bibitem{Banerjee2013} D. Banerjee, M. B\"ogli, M. Dalmonte, E. Rico, P. Stebler, U.-J. Wiese, and P. Zoller, Phys. Rev. Lett. {\bf 110}, 125303 (2013).
\bibitem{Zohar2013b} E. Zohar, J. I. Cirac, and B. Reznik, Phys. Rev. Lett. {\bf 110}, 125304 (2013).
\bibitem{Tagliacozzo2013} L. Tagliacozzo, A. Celi, P. Orland, M. W. Mitchell, and  M. Lewenstein, Nature Commun. {\bf 4}, 2615 (2013).

\bibitem{Zhang1989} S. C. Zhang, T. H. Hansson, and S. Kivelson, Phys. Rev. Lett. {\bf 62}, 82 (1989).
\bibitem{Rabello1996} S. J. Benetton Rabello, Phys. Rev. Lett {\bf 76}, 4007 (1996).
\bibitem{Fradkin1989} E. Fradkin, Phys. Rev. Lett. {\bf 63}, 322 (1989).
\bibitem{Aglietti1996} U. Aglietti L. Griguolo, R. Jackiw, S.-Y. Pi and D. Seminara, Phys. Rev. Lett.{\bf  77}, 4406 (1996).
\bibitem{Edmonds2013} M. Edmonds, M. Valiente, G. Juzeliunas, L. Santos, and P. \"Ohberg, Phys. Rev. Lett. {\bf 110}, 085301 (2013).

\bibitem{Keilmann2011} T. Keilmann, S. Lanzmich, I. McCulloch, and M. Roncaglia, Nature Commun. {\bf 2}, 361 (2011).
\bibitem{Greschner2015AHM} S. Greschner and L. Santos, arXiv:1501.07462.
\bibitem{Hao2009} Y. Hao, Y. Zhang, and S. Chen, Phys. Rev. A {\bf 79}, 043633 (2009).
\bibitem{DelCampo2008} A. del Campo, Phys. Rev. A {\bf 78}, 045602 (2008).
\bibitem{Hao2012} Y. Hao, and S. Chen, Phys. Rev. A {\bf 86}, 043631 (2012).
\bibitem{Wang2014} L. Wang, L. Wang, and Y. Zhang, arXiv:1411.5600.

\bibitem{MiyakeThesis} H. Miyake, Ph. D. thesis, Massachusetts Institute of Technology (2013).

\bibitem{Kardar86} M. Kardar, Phys. Rev. B 33, 3125 (1986)
\bibitem{Dhar2013} A. Dhar, T. Mishra, M. Maji, R. V. Pai, S. Mukerjee, and A. Paramekanti, Phys. Rev. B 87, 174501 (2013)
\bibitem{Petrescu2013} A. Petrescu and K. Le Hur, Phys. Rev. Lett. {\bf 111}, 150601 (2013). 
\bibitem{Tokuno2014} A. Tokuno, A. Georges, New J. Phys. 16, 073005 (2014)
\bibitem{Piraud2015} M. Piraud, F. Heidrich-Meisner, I. P. McCulloch, S. Greschner, T. Vekua, U. Schollw\"ock, arXiv:1409.7016
\bibitem{Greschner2015VL} S. Greschner, M. Piraud, F. Heidrich-Meisner, I. P.  McCulloch, U. Schollw\"ock, and T. Vekua, arXiv:1504.06564.
\bibitem{Schollwoeck11} U. Schollw\"ock, Annals of Physics {\bf 326}, 96 (2011).

\bibitem{Endres2011} M. Endres {\it et al.}, Science {\bf 334}, 200 (2011).
\bibitem{Greschner2014a} S. Greschner, L. Santos, and D. Poletti, Phys. Rev. Lett. {\bf 113}, 183002 (2014).
\bibitem{DallaTorre2006} E. G. Dalla Torre, E. Berg, and E. Altman, Phys. Rev. Lett. {\bf 97}, 260401 (2006).
\bibitem{Berg2008} E. Berg, E. G. Dalla Torre, Th. Giamarchi, and E. Altman, Phys. Rev. B {\bf 77}, 245119 (2008).
\bibitem{Greschner2013b} S. Greschner, L. Santos, and T. Vekua, Phys. Rev. A {\bf 87}, 033609 (2013).

\bibitem{Huerga2013} D. Huerga, J. Dukelsky, and G. E. Scuseria, Phys. Rev. Lett. {\bf 111}, 045701 (2013).  
\bibitem{Huerga2014} D. Huerga, J. Dukelsky, N. Laflorencie, and G. Ortiz, Phys. Rev. B {\bf 89}, 094401 (2014).

\bibitem{Bakr2010} W. Bakr {\it et al.}, Science {\bf 329}, 547 (2010).
\bibitem{Wuertz2009} P. Würtz, T. Langen, T. Gericke, A. Koglbauer, and H. Ott, Phys. Rev. Lett. {\bf 103}, 080404 (2009).
\bibitem{Cheneau2012} M. Cheneau {\it et al.}, Nature {\bf 481}, 484 (2012). 
\bibitem{Andraschko2015} F. Andraschko, and J. Sirker, Phys. Rev. B 91, 235132 (2015).

\end{thebibliography}

\end{document}